\documentclass[twocolumn,fp]{jpsj3}
\usepackage{cuted}
\usepackage{txfonts}
\usepackage{graphics}
\usepackage{dcolumn}
\usepackage{color}
\usepackage{array,multirow,makecell}
\usepackage{bm}
\usepackage{nicefrac}
\usepackage{multicol,blindtext}
\usepackage{amssymb}
\usepackage{kantlipsum,widetext}

\newcommand{\YRS}{YbRh$_2$Si$_2$ }

\title{Transport Spectroscopy of the Field Induced Cascade of Lifshitz Transitions in YbRh$_{2}$Si$_{2}$}

\author{Alexandre~Pourret$^{1}$\thanks{alexandre.pourret@cea.fr},
Sergei G.~Sharapov$^{2}$,
Tatsuma D.~Matsuda$^{3}$,
Georg~Knebel$^{1}$,
Gertrud~Zwicknagl$^{4}$,
Andrey A.~Varlamov$^{5}$
}

\inst{
$^1$Univ. Grenoble Alpes, CEA, IRIG, PHELIQS, F-38000 Grenoble, France\\
$^2$Bogolyubov Institute for Theoretical Physics, National Academy of Science of Ukraine, 14-b Metrologichna Street, Kiev, 03680, Ukraine\\
$^3$Advanced Science Research Center, Japan Atomic Energy Agency, Tokai, Ibaraki 319-1195, Japan\\
$^4$Institut f\"ur Mathematische Physik, Technische Universit\"at Braunschweig, Mendelssohnstr. 3, 38106 Braunschweig, Germany\\
$^5$CNR-SPIN, c/o DICII-Università di Roma Tor Vergata,  Via del Politechnico, 1, 00133 Roma, Italy\\
}

\abst{
A series of strong anomalies in the thermoelectric power is observed in the heavy fermion compound \YRS\ under the effect of magnetic field varying in the range from 9.5~T to 13~T. We identify these features  with a sequence of topological transformations of the sophisticated Fermi surface of this compound,  namely a cascade of Lifshitz topological transitions.  In order to undoubtedly attribute these anomalies to the specific topological changes of the Fermi surface, we employ the renormalized band method. Basing on its results we suggest a simplified model  consisting of the large peripheral Fermi surface sheet and the number of continuously appearing (disappearing) small "voids" or "necks".  We  account for the multiple electron scattering processes between various components of the Fermi surface, calculate the corresponding scattering times, and, finally, find the magnetic field dependence of the Seebeck coefficient. The obtained analytical expression reproduces reasonably  the observed positions of the maxima and minima as well as the overall line shapes  and allows us to identify the character of corresponding topological transformations.
}

\begin{document}
\maketitle

\section{Introduction}
The topology of Fermi surface (FS) is a fundamental property of metals and it is controlled by the periodicity and symmetry of the crystalline lattice and by occupation of the electronic energy bands. Under the influence of external effects such as chemical doping, pressure, or magnetic field ($H$), the Fermi energy or the band structure may change and a FS instability can be induced. If the number of  components of the FS changes under such impact, this alteration of the metal properties is called a Lifshitz transition \cite{Lifshitz1960, Varlamov1989, Blanter1994}.  Lifshitz transitions,  known since the 1960, are continuous quantum phase transitions at zero temperature.  Following the Ehrenfest terminology \cite{Ehrenfest1933}, they are referred to in the literature as $2{\frac{1}{2}}$ order phase transitions. At temperatures different from zero and/or in presence of impurities, such a FS reconstruction is often attributed to an electronic topological transition (ETT). Lifshitz transitions are associated with critical points in the energy bands, i. e., maxima, minima, and saddle points, whose presence follows directly from lattice periodicity. 

First experimental studies of Lifshitz transitions have been performed on thalium and its alloys under pressure \cite{Brandt1966}, and in  metallic alloys such as Bi$_{1-x}$Sb$_x$ \cite{Brandt1980}, or Mo$_{1-x}$Re$_x$ \cite{Velikodnyi1986} under the anisotropic deformation.  The changes in the FS topology in these experiments were registrated directly either by the Shubnikov-de Haas effect or by the de Haas-van Alfen effect.
The breakthrough in experimental studies of ETT was achieved by Egorov and Fedorov \cite{Egorov1982} performing  the chemical substitution in the alloy Li$_{1-x}$Mg$_x$. They discovered that its  thermoelectric power exhibits a large peak at $x=0.19$. The band structure calculations associated this point to the Lifshitz transition of the compound.  Such result was foreseen theoretically \cite{Vaks1981}  basing on simple differentiation of the kink in density of states in the vicinity of ETT in the  Mott's formula for the Seebeck coefficient. It has been recognized later that the situation is more complicated and the giant thermoelectric power peak close to ETT is due to the scatterings of the charge carriers from the whole peripheric area of the FS to its newborn part \cite{Varlamov1985}. 

Over the time it has been discovered that Lifshitz transitions appear in different classes of strongly correlated electron systems such as cuprate high-temperature superconductors \cite{LeBoeuf2011, Benhabib2015}, iron-based superconductors \cite{Liu2010, Malaeb2012}, etc.  More  recently the Fermi-surface instabilities have regained interest with the discovery of materials where the new exotic topological orders are expected \cite{Soluyanov2015, Liu2016}. It is worth to emphasize that the Lifshitz transitions are mainly induced by chemical doping in all mentioned above systems. Contrary, in extensively studied heavy-fermion systems the Lifshitz transitions are  driven by a magnetic field \cite{Daou2006, Harrison2007, Sebastian2009, Boukahil2014, Naren2013, Pfau2013, Pourret2013, Gourgout2016, Bastien2016, Bercx2012, Pfau2017, Ptok2017}. In these compounds, flat quasiparticle bands with the widths often comparable to the Zeeman splitting of the energy bands $(g_{\mathrm{eff} }/2)\mu_BH$ for field of order 10~T cross the Fermi level \cite{Thalmeier2005, Loehneysen2007}.  As a consequence, the Zeeman splitting can be strong enough to suppress the spin-split FS pockets leading to Lifshitz transitions \cite{Daou2006, Boukahil2014, Naren2013, Pfau2013, Pourret2013}. This scenario can explain the pseudo-metamagnetic transition in the paramagnetic CeRu$_2$Si$_2$ at $H_m \approx 7.8$~T.\cite{Daou2006}

In the present paper we will concentrate on the field induced Lifshitz transitions in \YRS. The latter is a well studied heavy fermion compound which orders antiferromagnetically at $T_N = 70$~mK. Thermodynamic measurements in applied magnetic field clearly show the existence of the two different energy scales. Firstly,  the antiferromagnetic order can be suppressed by applying the magnetic field of $H_c = 0.06$~T in the basal plane of the tetragonal crystal and $H_c =0.6$~T for $H \parallel c$. This field-induced quantum criticality has been intensively studied (see e.g. Ref.~\citen{Steglich2014} for a review). Recently it has been observed that \YRS   gets superconducting at the milli-Kelvin temperatures when the nuclear magnetic order.\cite{Schuberth2016} The experiments on \YRS  have played a key role in understanding of unconventional quantum criticality.\cite{Si2010}  Drastic changes in the Hall coefficient \cite{Paschen2004} and some thermodynamic properties \cite{Gegenwart2007} in the paramagnetic regime are observed at a crossoverline $T^\star (H)$. $T^\star$ extrapolates to the critical field $H_c$ at zero temperature. This jump in the Hall constant is discussed as a discontinuous change of the Fermi volume and as indication of a so-called Kondo-breakdown quantum critical point.\cite{Si2010} This scenario is based on the disintegration of the Kondo singlet at the critical field due to the antiferromagnetic order. However, experimentally these fluctuations between a small FS (in the antiferromagnetic state) and the large FS could not be observed by any direct measurement of the FS. Angular resolved photoemission spectroscopy, performed at temperatures above 1.5~K are in agreement with the so-called large FS,\cite{Wigger2007,Kummer2015} what may be too high to identify the FS at $T=0$. \cite{Paschen2015}An alternative explanation for the anomalies along the $T^\star (H)$-line has been suggested by Abrahams, Schmalian and Wölfle in a series of
papers. \cite{Abrahams14,Woelfle17,Woelfle18} The central assumption of this ``critical quasiparticle'' scenario is that the quasiparticles remain robust but are modified by scattering from critical spin fluctuations. An important consequence is that the volume of the Fermi surface, i. e., the number of itinerant fermions, does not change across $T^\star (H)$. The results of this theory are in good agreement with experimental data. \cite{Wetli18, Pal19}
\begin{figure}[th]
\begin{center}
\includegraphics[width= 0.9\columnwidth]{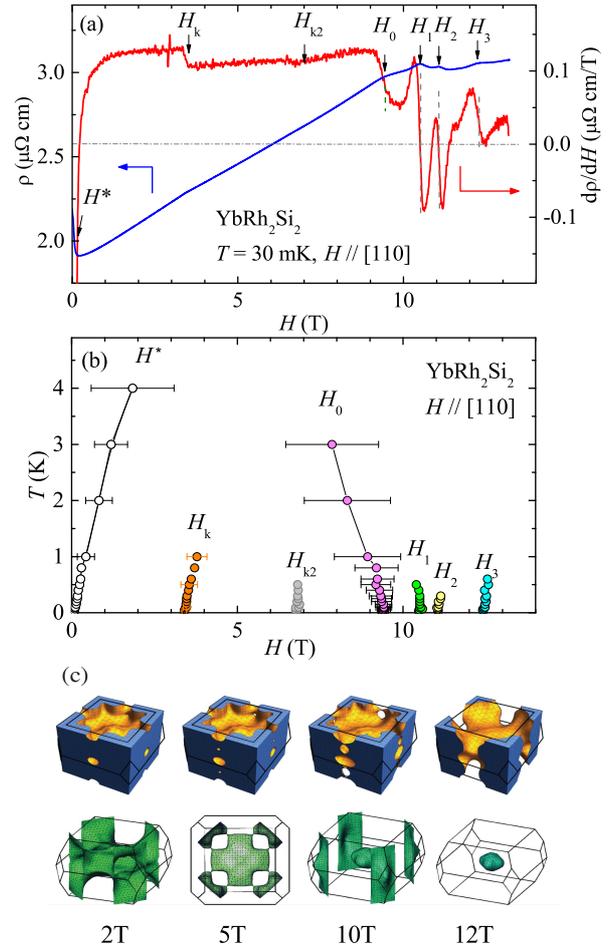}
\end{center}
\caption{(Colore online) (a) Magnetoresistance of \YRS at 30~mK (see Ref.~\citen{Pourret2013}) and its derivative over temperature. (b) Magnetic field temperature phase diagram from the magnetoresistance at different temperatures. $H_0$ corresponds to the field where the main Lifshitz transition occurs, i.e. the field where the van Hove singularity of the minority spin density of state crosses the Fermi level. Above $H_0$, successive topological changes ($H_1$, $H_2$, $H_3$) occur on both minority FSs. (c) Field evolution of the minority FS pockets: doughnut FS in the electron representation (upper row), jungle-gym FS (lower row) taken from Ref.~\citen{Pfau2013}.}
\label{fig01}
\end{figure}

A second characteristic field appears at $H_0 \approx10$~T applied in the $ab$ plane. Here the magnetization $M(H)$ shows a strong  decrease of the slope\cite{Gegenwart2006}, with no saturation up to 35~T. There is  no indication of a metamagnetic transition or any field induced magnetic phases around $H_0$. This change in the magnetization is  accompanied by a strong decrease of the Sommerfeld coefficient of the specific heat, the coefficient in front of $T^2$ in the electrical resistivity temperature dependence, and the linear magnetostriction coefficient,\cite{Gegenwart2006, Knebel2006}. All these facts  indicate on  a strong decrease of the density of states. Quantum oscillations experiments indicated that the observed oscillation frequencies have a strong field dependence through the field of 10~T, and these changes have been discussed as indication for a Lifshitz transition near $H_0$.  Thermoelectric power experiments\cite{Pfau2013, Pourret2013} and also magnetotransport\cite{Naren2013} showed a complex phase diagram with several anomalies which has been interpreted as Lifshitz transitions. In Fig.~\ref{fig01}(a) we show the magnetoresistance and its derivative $d\rho / dH$ at $T=30$~mK up to 13~T showing clearly these anomalies which allows to determine the phase diagram [see Fig.~\ref{fig01}(b)]. 
The interpretation of successive Lifshitz transitions has been supported by renormalized band structure calculations. These calculations show that the Lifshitz transitions are the result of the field evolution of the spin-split density of states.\cite{Pfau2013, Naren2013} At zero field, the density of states shows one pronounced and several weaker van Hove singularities slightly below the Fermi energy $E_F$ \cite{Naren2013}. There are also van Hove singularities in the empty part above $E_F$ but they appear at higher energies. Under magnetic field the energy bands split in majority and minority subbands.  While the majority spin subband shifts to lower energies, the Fermi level does not cross a van Hove singularity for the magnetic fields under consideration. The smooth evolution with magnetic field of the majority Fermi surface is in marked contrast to the behavior of its minority counterpart. As the minority spin subbands are shifted to higher energies the Fermi level
crosses the above mentioned van Hove singularities. Three topological transitions have  been  identified, and most importantly near $H_0\approx$9.5~T the van Hove singularity of minority spin density of state crosses the Fermi level giving rise to a strong reduction of the density of states. The field evolution of the minority FS pockets is shown in Fig.~\ref{fig01}(c). The upper row shows the field evolution of the minority spin band corresponding to the doughnut in the limit of vanishing field \cite{Naren2013}. The main Lifshitz transition corresponds in this presentation to the connection at the hole at the f and s point at the Brillouin zone at $H_0$. The lower row shows the field evolution of the minority spin band corresponding to the jungle gym in the limit of vanishing field (taken from Ref.~\citen{Pfau2013}). The cascade of topological changes, discussed in Refs.~\citen{Pfau2013, Pourret2013}, comes from the collapse of the corrugated cylinder above $H_0$, between 10~T and 12~T. 

In the following, we will concentrate on the cascade of Lifshitz transitions above $H_0$ in YbRh$_2$Si$_2$ to get the precise shape of the different anomalies. 
Renormalized band (RB) structure calculations \cite{Zwicknagl2011, Andersen1975} describing in details the effect of a magnetic field on the quasiparticle bands serve to identify "regular" regions in $k$-space where the FSs evolve continuously with magnetic field and "critical" regions surrounding the critical points where Lifshitz transitions may occur. A new approach to calculate of the Seebeck coefficient near the critical regions taking into account interference phenomena in the  scattering process is presented. We will discuss in detail, how the pillow FS of the minority jungle-gym (Fig~\ref{fig01}(c) lower row) will vanish by three distinct topological transitions.
We find that the calculated magnetic field dependence of the Seebeck coefficient turns to be in good agreement to the experimental findings, allowing a real transport spectroscopy of  FS topological transitions.

\section{Thermoelectric power measurements}

\begin{figure}[th]
	\includegraphics[width=\columnwidth]{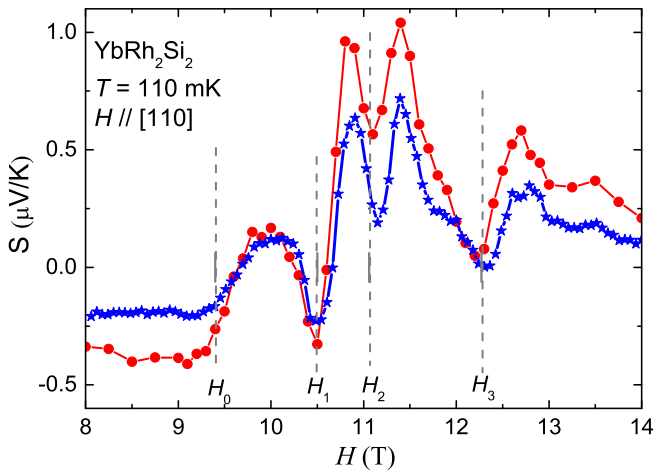}
	\caption{(Color online) The magnetic field dependence of the Seebeck coefficient between 8~T and 14~T in the heavy-fermion compound YbRh$_2$Si$_2$ at 110~mK obtained from two different cooling on the same sample. The critical fields $H_1=10.5$~T, $H_2=11.1~$T, and $H_3=12.2$~T correspond to the position of the series of minima.}
	\label{fig02}
\end{figure}

We remeasured the thermoelectric power on the same wisker-like single crystal of \YRS\ as previously in narrower steps. \cite{Pourret2013} The magnetic field is applied along [110] of the tetragonal crystal and heat or charge current perpendicular to the field along $[1\bar{1}0]$, which corresponds to the long axis of the crystal. The thermoelectric power has been measured by a one heater, two thermometers set-up in a dilution refrigerator down to 110~mK and up to 16~T. Thermometers and heater are thermally decoupled from the sample holders by highly resistive manganin wires.  Figure \ref{fig02} shows the field dependence of the Seebeck coefficient at 110~mK for both experiments. 

\begin{figure}[h]
\includegraphics[width=0.85\linewidth]{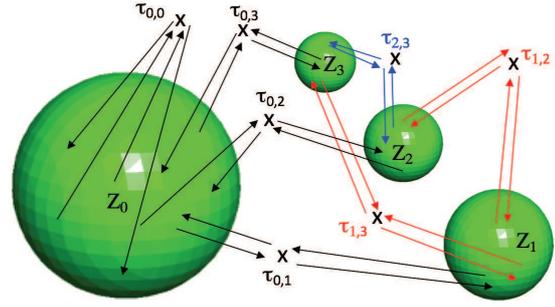}
\caption{(Color online) Schematic representation of the various quasiparticle scattering processes between the pockets with the energy sizes $Z_{i}$. $\tau_{i,j}$ correspond to the processes from pocket $i$ to pocket $j$ ($\tau_{0,j}$ refers to the process from the main Fermi surface $Z_{0}$)  through impurities (crosses).}
\label{fig03}
\end{figure}
\section{Analysis of the thermoelectric power}
\subsection{Cascade of Lifshitz transition and quasiparticle scattering}

To take into account the specificity of a cascade of Lifshitz transitions on quasiparticle scattering,
we firstly assume that all topological changes of the FS are due to the formation of new spherical pockets (see Fig. \ref{fig03}). Here the index $i=0$ denotes to regular part of the FS hereafter referred to as mainland while the critical parts are labelled by $i=1,...N$. The quasiparticle excitation spectrum for each of them can be presented as:
\begin{equation}
\xi _{i}(\mathbf{p})=\frac{(\mathbf{p}-\mathbf{p}_{ci})^{2}}{2m_{i}}%
-Z_{i},\qquad i=0,\ldots N,  \label{spectra}
\end{equation}
where $\mathbf{p}_{ci}$ is the position of its center in the Brillouin zone, $Z_{i}=\mu -\epsilon _{ci}$ is its energetic size, $\epsilon_{ci}\left( H_{i}\right) =\mu $ is the critical point of multi-valued function of energy when the i-th voids appears, $m_{i}$ is the corresponding effective mass.
In Ref. \citen{Varlamov1985, Varlamov1988} was demonstrated, that the anomaly in the  Seebeck coefficient in the vicinity of Lifshitz transition is directly related to the specific quasiparticle scattering starting from the bulk component of the mainland FS (the energy size is $Z_{0}$) and ending at the small pocket of the FS  (characterized by the energy size $Z_{i}$). It was shown that the latter is a trap for quasiparticles,  their velocity here is small and they die away. The next scattering with the dominating probability returns the quasiparticle back to the mainland FS. Such forth and back scattering give rise to the singularity of the Seebeck coefficient. At the first glance generalization of the described scattering mechanism to the case of multiple Lifshitz transitions seems to be trivial: it is just necessary to account for the round trips of the quasiparticles from the mainland FS to all other pockets ($Z_{1}, Z_{2}, Z_{3} \ldots $), see $\tau_{0,j}$ processes in Fig. \ref{fig03}. Yet, as will be shown below, this is not enough. An important role is also played by the ``traveling'' of the quasiparticles between the newborn FS elements, $\tau_{1,j}$ and $\tau_{2,j}$ processes in Fig. \ref{fig03}.
The expression for the scattering time for the quasiparticles belonging to the ``$l$''-th pocket accounting for the return to the same pocket and round trips to the smaller ones with $i>l$ (see Fig. 1), in full analogy with the results of Ref. \citen{Varlamov1985, Varlamov1988}, is given by:\\
\begin{equation}
\tau^{-1} _{l}\left( \omega ,Z_{l},..Z_{N}\right) =\sum_{i=l}^{N}{\tau }
_{l,i}^{-1}(\omega ,Z_{l}),
 \label{taul}
\end{equation}
where
\begin{equation}
\tau_{l,i}^{-1}(\omega ,Z_{l})=\frac{\kappa _{l}\left(
	-Z_{i}-\omega \right) }{2{\tilde\tau}_{l}\sqrt{Z_{l}}}.
\label{taui}
\end{equation}
Here the function
\begin{equation}
\kappa _{l}(\varepsilon )=\sqrt{2}\sqrt{(\varepsilon ^{2}+\xi Z_{l})^{1/2}-\varepsilon},
 \label{kl}
\end{equation}
the parameter ${\tilde{\tau}}_{l}^{-1}=\pi ^{-1}n_{\mathrm{imp} }|U|^{2}m^{3/2}(2Z_{l})^{1/2}=2\sqrt{\xi Z_{l}}$, $\xi=n^2_{imp}U^4m^3/(2\pi)$,  $\omega $ is the energy, $n_{\mathrm{imp}}$ is the impurities concentration, and $U$ is the amplitude of the scattering potential, which we assume independent of momentum. For the sake of simplicity we  assume the effective masses for electrons in all voids to be equal, $m_{i}=m$.
\begin{figure}[th]
	\includegraphics[width=\columnwidth]{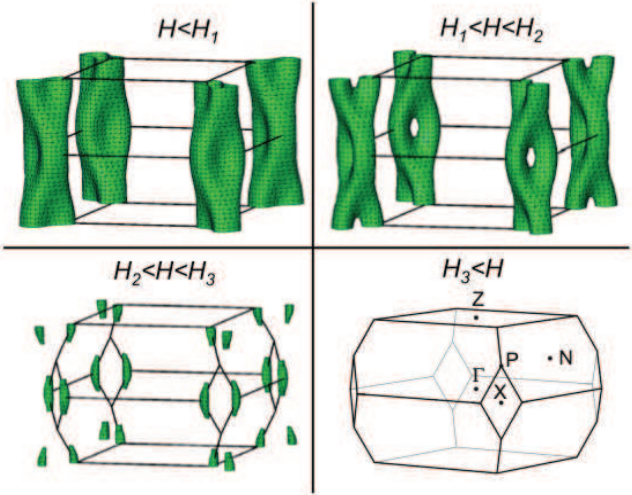}
	\caption{(Color online) Evolution with magnetic field of the critical part of the minority jungle gym sheet above $H_0$. There is a series of Lifshitz transitions which is seen in the evolution of the FS. At $H_1$, apertures opening in the cylinders, followed at $H_2$ by a neck breaking of the cylinders and finally at $H_3$ the remaining ellipsoids collapse.}
	\label{fig04}
\end{figure}
\subsection{Renormalized band calculation of the Fermi surface}

In order to model the specific case of a cascade of Lifshitz transitions in YbRh$_2$Si$_2$, we need to know precisely the evolution of the FS with magnetic field.  The cascade of Lifshitz transitions  give rise to series of anomalies in the Seebeck coefficient above $H_0=9.5$, see Fig. \ref{fig02} \cite{Pourret2013}.  The corresponding behaviour of the resistivity is shown in Fig. \ref{fig01}. One can see that the variation of resistivity in the whole range of considered magnetic fields is minut and the position of anomalies can be determined either from its derivative or, better, from the measured Seebeck signal. 
We will concentrate our study on the main anomalies: the sharp minima at  $H_1=10.5$~T and $H_2=11.1$~T, and  $H_3=12.2$~T.  When the temperature increases, the different anomalies broaden and merge together into one large transition at $H_0$. In order to relate the different anomalies observed in the Seebeck coefficient to the topological changes occurring in the FS, the RB scheme, which combines material-specific ab-initio methods and phenomenological considerations in the spirit of the Landau theory of Fermi liquids has been used \cite{Zwicknagl2016}. The band-structures were obtained by the fully relativistic formulation of the linear muffin-tin orbitals (LMTO) method \cite{Andersen1975, Skriver1984, Christensen1984}. The progressive de-renormalization of the quasiparticles with increasing magnetic field and the correlation-enhanced Zeeman splitting were calculated from the Renormalization Perturbation Theory (RPT) \cite{Edwards2011, Edwards2013}.

For $H=0$, there are three bands intersecting the Fermi energy, $E_F$. In the following discussion, we shall neglect the small $\Gamma$-centered electron pocket and focus on the two bands (35 and 37) giving rise to the two major sheets. Under magnetic field, these two bands will give rise to spin splitted FS pocket, only minority spin bands become critical under magnetic field as shown on Fig.\ref{fig01}.  The shape of the FS agrees with ARPES measurements \cite{Kummer2015} and the overall topology qualitatively agrees with local density approximation (LDA) results \cite{Knebel2006, Jeong2006, Rourke2008, Rourke2009, Sutton2010}. The dominant contribution to the quasiparticle density of states comes from the $Z$-centered peripherical FS which has predominantly $f$-character while the states forming the {}``jungle gym'' are strongly hybridized.

Yb heavy-fermion systems can be considered as hole analogues of Ce counterparts. As a consequence, the center of gravity of the heavy quasiparticle bands is below the Fermi energy and the Zeeman splitting of the FSs will strongly affect the minority spin FS \cite{Zwicknagl2016, Zwicknagl2011, Zwicknagl2011b, Pfau2013}. As shown in Fig. \ref{fig04}, the corrugated cylinders of the minority jungle gym surface are expected to disappear in the narrow magnetic field range 10-13~T (just above $H_0$=9.5T) while the other surfaces hardly change \cite{Pfau2013}.
At $H_1$, the apertures open at the $X$ points, changing the connectivity of the FS: one can consider this transformation of the FS as the electron voids formation. At $H_2$ the FS topology changes again thinning cylinders at this field break to ellipsoids centered at the $Y$ points of the Brillouin zone. Finally, at $H_3$ the electronic ellipsoids disappear at all.
The above numerical findings indicate on the sequence and the character of Lifshitz transitions occurring in YbRh$_2$Si$_2$, viz. the formation of electron void followed by the neck disruption and  all these transformations are ended by the electron ellipsoid collapse.
Here it is worth to mention that the Eq. (\ref{taul}) for relaxation time corresponds to the case of  above considered void formation, i.e. it is applicable to the first and the third transitions of the considered sequence.  Yet, as it was demonstrated in Refs. \cite{Varlamov1988, Varlamov1989},  the relaxation times for both types of topological changes of the FS in three dimensional case are very similar. Namely, there is a simple symmetry transformation $Z_{\mathrm{sp}}\rightarrow -Z_{\mathrm{hp}};\omega_{\mathrm{sp}}\rightarrow -\omega _{\mathrm{hp}}$ $\kappa_{sp}\rightarrow - \kappa_{hp}$ ("$\mathrm{hp}$'' and ``$\mathrm{sp}$'' are the labels corresponding to the neck disruption and void fromation transition respectively \cite{Varlamov1989}.
\begin{figure}[h!]
	\includegraphics[width=0.8\columnwidth]{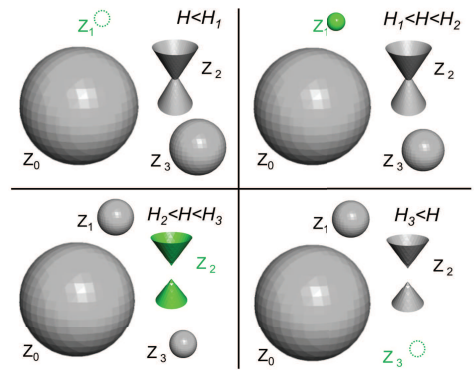}
\caption{(Color online) The scenario of the multiple Lifshitz transitions evolution: at $H_1$ the pockets $Z_1$ appears, at $H_2$ a neck breaking occurs in the pocket $Z_2$, finally at $H_3$ the pocket $Z_3$ disappears.}
\label{fig05}
\end{figure}
The characteristic distance between two consequent LTs coming from the calculation in the cascade is of the order of $0.5$~T, which is close to the distance between peaks in the Seebeck signal, see Fig. \ref{fig02}. Note that due to the large $g$-factor the scale of $0.5$~T corresponds to the energy $\sim 10$~K.
\subsection{Toy model of the FS and Seebeck coefficient for YbRh$_2$Si$_2$}



In order to adjust the above considerations on the multiple Lifshitz transitions to the RB calculations in the specific case of YbRh$_2$Si$_2$ we assume the existence of four sheets of the FS in the Brillouin zone. Starting with a mainland part $Z_0$ which practically remains unchanged in the studied range of magnetic fields (11-13~T), we need three apparitions (disparition) of pockets/neck to model the evolution of the corrugated cylinders of the minority spin FS, see Fig. \ref{fig05}. Below the field $H_1$ the FS consists of the mainland, the relatively large pocket  $Z_3$ and the sheet of FS close to neck disruption Lifshitz transition $Z_2$.  The  small pocket $Z_1$ appears when the magnetic field reaches $H_1$, corresponding to $\epsilon_{c1}(H_1)=\mu$. In others words this topological transition occurs when the energy size of the small pocket $Z_1(H)=\mu-\epsilon_{c1} (H)=\epsilon_{c1}(H_1)-\epsilon_{c1}(H)$ becomes positive ( $\epsilon_{c1}(H)$ is a decreasing function). The apparition of this new pocket $Z_1$ is equivalent to the opening of apertures in the cylinders of the minority spin band, see Fig. \ref{fig04}. With further increase of the magnetic field, a neck breaking of the sheet $Z_2$ occurs at $H_2$. This transition models the evolution of the FS from the cylinders to the small ellipsoids ($Z_2(H)=\epsilon_{c2}(H)-\mu = \epsilon_{c2}(H) -\epsilon_{c2}(H_2)$ where  $\epsilon_{c2}(H)$ is a decreasing function). Similarly, the collapse of  the last pocket $Z_3$ at $H_3$ (with $Z_3(H)=\mu -\epsilon_{c3}(H) = \epsilon_{c3}(H_3)-\epsilon_{c3}(H) $, $\epsilon_{c3}(H)$ is a increasing function) symbolizes the complete disappearance of the ellipsoids of minority spin band, see Fig.~\ref{fig04}. The nomenclature of the toy model is summarised in the Table \ref{Table}, where the different energy parameter $Z_i(H)$ are defined with their dependence in field and the associated range of fields where they are valid. The thermoelectric coefficient $\beta$ which determines the part of electric current related to the temperature gradient applied to the system ($j^{(h)}_{e}=-\beta \nabla T$) can be expressed in terms of the imaginary part of the one-electron Green function. In the case of multi-sheet FS it takes the form \cite{Varlamov1989}:
\begin{equation}
\beta =-\frac{e}{2\pi T^{2}}\int \frac{\omega d\omega }{\cosh ^{2}\left(\hbar \omega /2k_{B}T\right) }\sum_{l=0}^{N}\int v_{l}^{2}\frac{d^{3}p}{\left( 2\pi\right) ^{3}}\mathrm{Im}^{2}G_{ll}^{R}
\end{equation}
with $G_{ll}^{R}$ as the retarded Green function of quasiparticle at the $l$'s sheet of FS, with corresponding scattering time (\ref{taul}), and  $v_{l}$ as velocity. Finally, the Seebeck coefficient $S$ can be obtained from the thermoelectric one using the relation:
$S = - \beta/\sigma$, where $\sigma$ is the diagonal electrical conductivity of the sample. The latter, as one can see from the Fig. \ref{fig01}, does not undergo considerable changes in the vicinity of Lifshitz transition (see \cite{Varlamov1989, Pourret2013}) and we assume it as constant:
\begin{widetext}
\begin{eqnarray}
S_{xx}(H) =
\frac{k_B}{e}\int_{-\infty}^{\infty} \frac{x d x}{\cosh^2 x}
\left[\frac{2\tilde \sigma_0  \sqrt{Z_0}}
{\kappa _{0}(-Z_{0})+\kappa _{0}(- 2 x T
	-Z_{1})-\kappa _{0}(2 xT +Z_{2})+\kappa _{0}(- 2 xT -Z_{3})}\right.
\nonumber \\
\left. +
\frac{2\tilde \sigma_2  \sqrt{|Z_2|}}{\kappa_2(Z_2) + \kappa _{2}(-2xT -Z_{1})}+
\frac{2 \tilde \sigma_3 \sqrt{Z_3} \theta (Z_{3})}{%
	\kappa_3 (-Z_{3})-\kappa _{3}(2xT +Z_{2})+\kappa _{3}(-2xT -Z_{1})}\right] .
\label{betalast}
\end{eqnarray}
\end{widetext}

\begin{table*}[h!]
\begin{tabular}{|c|c|c|c|c|}
\hline 
FS sheet & Dependence on $H$  & Range of fields,  & Type of  & \textquotedblleft Toy\textquotedblright{} depen- 
\\
&  & where FS sheet exists  & singularity  & dence on $z\propto H$ \\ 
\hline
\hline
Mainland $Z_{0}$  & const  & Any field  & Nonsingular  & $Z_{0} = \mbox{const} $  %
\\
&  &  & periphery of FS & \\
\hline 
``small'' pocket $Z_{1}$  & $Z_{1}(H) =\mu(H_{1})-\epsilon_{c1}(H) $ & $H > H_{1}$  & Void  & $Z_{1} =Z_{10}-z$ 
\\
& $= \epsilon_{c1}(H_{1})-\epsilon_{c1}(H)$  &  & (apparition at $H_{1})$ &
 \\
 \hline
Neck $Z_{2}$  & $Z_{2}(H) = \mu(H_{2})-\epsilon_{c2}(H_2)$ & Any field FS sheet  & Neck & $Z_{2} = Z_{20} + z$
\\
&  $=\epsilon_{c2}(H)-\epsilon_{c2}(H_{2})$  & changes its connectivity at $H_{2}$  & (disruption at $H_{2}$) &
\\
\hline 
``large'' pocket $Z_{3}$  & $Z_{3}(H) =\mu(H_{3})-\epsilon_{c3}(H)$ & $H < H_{3}$  & Void  & $Z_{3} = Z_{30} + z $
\\
& =$\epsilon_{c3}(H_{3})-\epsilon_{c3}(H)$ &  & (disparition at $H_{3}$) & \\
\hline 
\hline
\end{tabular}
\caption{The nomenclature of the toy model describing the multiple Lifshitz transitions evolution.}
\label{Table}
\end{table*}

Here $\tilde \sigma_l$ is the relative contribution of the $l$-th sheet to the total conductivity, $\theta(x)$ is the Heaviside step-function, $k_B$ is the Boltzmann constant, and $e$ is the absolute value of the electron charge. 

\begin{figure}[h!]
	\includegraphics[width=0.92\columnwidth]{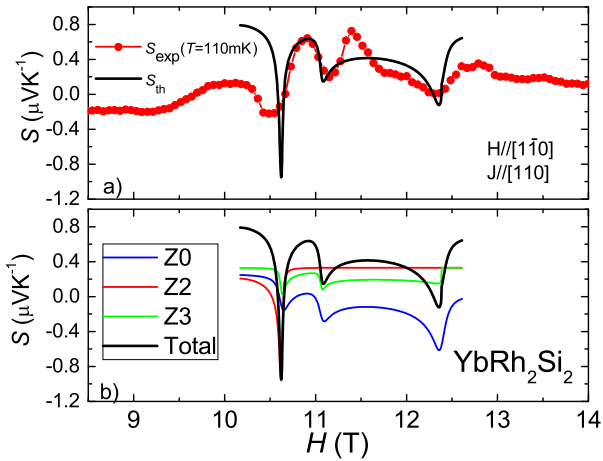}
\caption{(Color online) (a) Magnetic field dependence of the Seebeck coefficient through a cascade of Lifshitz transitions obtained at 110~mK compared with the theoretical prediction based on multi-sheet FS scattering processes. (b) Theoretical modeling of the contributions corresponding to quasiparticle  scatterings from the different sheets of the FS to the Seebeck coefficient with $\tilde \sigma_0=0.4$, $\tilde \sigma_2=0.01$, and $\tilde \sigma_3=0.59$ in Eq. (\ref{betalast}). }
\label{fig06}
\end{figure}

Three terms in Eq. (\ref{betalast}) correspond to the  inverse relaxation times of the quasiparticles traveling in the processes of elastic scatterings between the different sheets of FS. The first term accounts for the carriers scatterings from the ``mainland'' $ Z_0$ to itself and to the ``pockets'' $ Z_1, Z_2, Z_3$. The second term  in Eq. (\ref{betalast}) corresponds to both, the intra-pocket and inter-pocket scattering processes which transfer the carriers from the  pocket with the conic singularity $Z_2$  to the spherical pocket $Z_1$. Note, that the functions $\kappa _{i}(2xT+Z_2)$, corresponding to the scattering processes involving  the neck, enter with the negative sign (see above). Finally, the third term in Eq. (\ref{betalast}), besides the intra-pocket scatterings of carriers describes the inter-pocket ones  from the electron pocket $Z_3$ to $  Z_1, Z_2$. The remaining intra-pocket scattering processes involving $Z_1$ are neglected in comparison to the other terms in view of the smallness of the corresponding density of states. Let us note that the inter-pocket scattering processes involving the pocket $Z_1$ and other pieces of the FS are already accounted for in Eq. (\ref{betalast}) by the first three terms.
Eq. (\ref{betalast}) formally determines the dependence of the Seebeck coefficient on magnetic field in the case of cascade Lifshitz transitions under consideration. Unfortunately, we do not have analytical expressions for the branches of the multi-valued function of energy  $\epsilon_{ci}(H)$. This is why, in purpose to demonstrate how Eq. (\ref{betalast}) works, we use the toy model dependence of $Z_i$:
$ Z_0 = \mbox{const}, \, Z_1 = (Z_{10} - z) , \, Z_2 =(Z_{20} + z), \, Z_3 =(Z_{30} + z)$, and study
$S(z)$ as the function of a single control parameter $z$. In view of relatively narrow interval of magnetic fields one can believe that $z \varpropto H$.
and $\mu \gg |Z_1|, |Z_2|, |Z_3|$,  $Z_{10}>0$, $Z_{30} < Z_{2 0} < 0$ are the constants.
In Fig. \ref{fig06}(a), the observed Seebeck signal at 110~mK and the corresponding theoretical prediction based on Eq. (\ref{betalast}) are represented.  One can see that the very complicated structure of experimentally observed maxima and minima is reasonably reproduced even in the framework of the chosen simple toy model of the spectrum.The only exception is the peak close to the field $H^*$= 11.4 T.    No topological changes of the FS was found by ab initio calculus here and, correspondingly, this explicitly observable singularity in Seebeck coefficient is not reflected by our toy model. 
 In  Fig. \ref{fig06}b) the partial contributions to the Seebeck coefficient determined by the scattering  of quasiparticles between the different sheets of the FS are represented. One can see that the above mentioned minimum is due to the scattering of the quasiparticles from the neck domain $Z_2$ to that one where the electron pocket $Z_1$ should occur when the magnetic field will reach the value $H_1$. This is the consequence of the well known in quantum mechanics scattering from the virtual level.  The abrupt fall of the contribution corresponding to the pocket $Z_3$ is the artifact caused by the simplified Theta-function modeling, which excludes the scattering at the virtual level.
 
\section{Conclusions}
We demonstrated that the Seebeck coefficient is very sensitive to topological changes of the FS in YbRh$_2$Si$_2$.  Using renormalization perturbation theory, the effect of magnetic field on the topology of the FS is precisely described, in particular the complex changes occurring in the minority spin FS from corrugated cylinders to ellipsoids. Using a simple model of continuously appearing (disappearing) pockets of the FS, the four main anomalies of the Seebeck coefficient can be well described. We have demonstrated that measurements of the Seebeck coefficient may provide very detailed information on the electronic structure in the vicinity of critical points and  that they may allow  for transport spectroscopy when only part of the FS becomes critical.
Let us underline the difference between the developed above consideration of the cascade Lifshitz transition in their complexity of scattering processes (see Eq. (\ref{betalast}))  from the theoretical treatment of the experimental findings in the ferromagnetic Kondo lattice material YbNi$_4$P$_2$ \cite{Pfau2017}.   In the latter work the behaviour of the Seebeck coefficient as a function of magnetic field  is described  in terms of the set of isolated independent Lifshitz transitions \cite{Varlamov1989, Blanter1994}. Each peak is interpreted as a neck disruption or void creation for electron or hole-like carriers. This simple approach  assumes the quasiparticle exchange between some fictitious, changing for each transition, mainland $\tilde{Z_0}$ and newborn at the appropriate magnetic field element of the FS  $\tilde{Z_i}$. In accordance to the standard theory of Lifshitz transitions the Seebeck coefficient close to the  anomaly is described by the expression:

\begin{equation}
S_{xx}(H) \!= \!
\frac{k_B}{e} \int_{-\infty}^{\infty}  \frac{x d x}{\cosh^2 x}
\frac{2\tilde \sigma_0  \sqrt{\tilde{Z_0}} }{ \kappa _{0}(-\tilde{Z_0}) \pm \kappa _{0}(\mp 2 x T \mp \tilde{Z_i}) } 
\label{single}.
\end{equation}
Accordingly, each anomaly treated in this way possesses the characteristic asymmetric shape and its overall sign reflects electron/hole type of carriers. Namely such situation corresponds to the experimental findings of the Ref.   \cite{Pfau2017} . It is easy to see, that Eq. (\ref{single}) follows from the first term of our general Eq. (\ref{betalast}) considered in the vicinity of each separate transition.
Let us underline, that the superposition of the individual singularities occurring due to the different terms of the Eq. (\ref{betalast}) can distort the  characteristic feature of the Lifshitz transition, the asymmetry of the Seebeck signal anomaly, as it takes place with the negative spike close to the field $H_1$ (see Fig. \ref{fig06}).
The theoretical analysis  of Ref. \citen{Pfau2017}  is valid for standing apart peaks, viz. when the width of the peak is much smaller than the distance between them.  When the  peaks are close to each other our multiple scattering approach accounting for the superposition phenomena between the new pockets occurring in the narrow energy range seems to be more relevant.  Comparing the experimental situations in YbRh$_2$Si$_2$ and that one of Ref.  \cite{Pfau2017}  one can see that the minimal distance between the anomalies in the latter is 0.75 T (for the most of the peaks it is a few Tesla), while all  anomalies in our experiments are separated by the value of 0.6 T.

\section*{Acknowledgments}	
Two of the  authors, S.~G. Sh. and A.~A.~V., acknowledge the support of European Union's Horizon 2020 research and innovation program under the grant agreements No. 644076 (CoExAn) and No. 731976 (MAGENTA). G.~Z. acknowledges financial supports of Univ. Grenoble Alpes and LPMMC for her stays in Grenoble and the joint French-German ANR-DFG grant Fermi-NESt. T.~M. acknowledges JSPS KAKENHI JP15H05884 grant. S.G.Sh. acknowledges the support by the National Academy of Sciences of Ukraine (project No. 0117U000236).

\end{document}